\documentclass[aps,prl,twocolumn,showkeys,showpacs,superscriptaddress,groupedaddress]{revtex4}  

\usepackage{graphicx}  
\usepackage{dcolumn}   
\usepackage{bm}        
\usepackage{amssymb}   
\usepackage{latexsym,bm}
\usepackage[tbtags]{amsmath}

\begin{document}

\widetext



\title{Singularity-free approximate analytical solution of capillary rise dynamics}

\author{Bohua Sun}

\affiliation{%
Cape Peninsula University of Technology, Cape Town, South Africa\\
sunb@cput.ac.za
}%



\begin{abstract}
\small
Capillary rise is one of the most well-known capillarity; however, no single and complete analytic solution has ever been obtained yet. This paper used the singularity-free equation, and successfully obtained its Taylor's series solution. The solution revealed that capillary rise dynamics is mainly controlled by the Bond number and the Galileo number, while the Bond number is a key parameter within the solution. To avoid the poor rate of convergence of Taylor's series solution, an approximate analytic solution was proposed, which was verified numerically.
\end{abstract}

\maketitle

\begin{figure}[h]
\centerline{\includegraphics[scale=0.7]{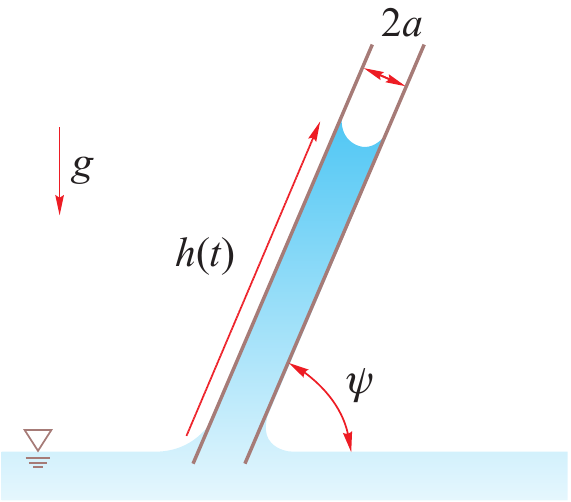}}
\centerline{\includegraphics[scale=0.7]{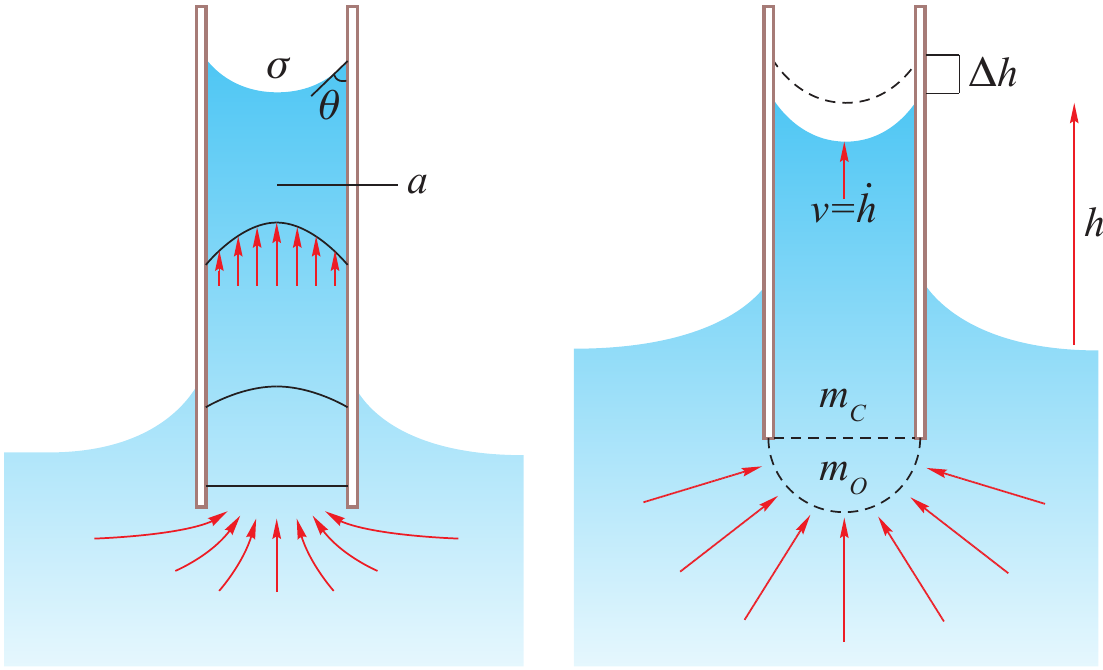}}
\caption{\label{fig1} The dynamics of capillary rise, capillary hight $h(t)$, surface tension $\sigma$, capillary tube radius $a$, wetting angle $\theta$, gravity $g$, inclined angle $\psi$}
\end{figure}

Capillary rise is one of the most well-known and vivid illustrations of capillarity (shown in Figure \ref{fig1}). Knowledge of capillarity laws is important in oil recovery, civil engineering, dyeing of textile fabrics, ink printing, and a variety of other fields. It is capillarity that brings water to the upper layer of soils, drives sap in plants, or lays the basis for the operation of pens \cite{was,zhm,bos,bri,sze,mag,zha2,bus,sun1,zho}.

Washburn \cite{was} developed an equation to describe the rate of liquid penetration into small cylindrical capillaries based on the Poiseuille flow profile, when neglecting the air resistance, a commonly used form of Lucas-Washburn's equation is $8\mu h\frac{dh}{dt}+\rho g h a^2 \sin \psi=2 a \sigma \cos \theta$. When the gravitational force is negligible, a well-known Washburn's law can be obtained with the initial condition $h(0)=0$: $h=\sqrt{\frac{a \sigma \cos \theta}{2 \mu}t}$, which predicts burst-like behavior with the velocity being infinite at zero time. This singularity of the solution highlights a deep inconsistency of the above equation \cite{zhm}. Taking into account the liquid's momentum in the tube and end-effect drag on the fluid entering the tube, Brittin \cite{bri} derived a more rigorous formulation of the Lucas-Washburn equation, as follows: $h\frac{d^2h}{dt^2}+\frac{5}{4}(\frac{dh}{dt})^2+\frac{8}{\rho a^2}\mu h \frac{dh}{dt}+ g h=\frac{2}{\rho a}\sigma \cos \theta$, where $\mu$ is the viscosity. The most popular equation is its cosmetic format $h\frac{d^2h}{dt^2}+(\frac{dh}{dt})^2+\frac{8}{\rho a^2}\mu h \frac{dh}{dt}+ g h=\frac{2}{\rho a}\sigma \cos \theta$, however, it has a singularity of $t=0$, namely $\dot{h}=\frac{dh}{dt}\rightarrow \infty$ as $t\rightarrow 0$, which would lead to an ill-posed problem. This equation cannot even deal with the natural initial conditions $h(0)=\dot{h}(0)=0$. Hence, formal remedy has to be taken $\dot{h}(0)=\sqrt{2\sigma \cos \theta /(\rho a)}$ \cite{bos}, neglecting such a logical drawback as the acceleration of the liquid front at zero time, which is infinite.

This singularity problem was first pointed out by Szekely \emph{et al}.\cite{sze}, who successfully removed the singularity problem by composing the correct energy balance for the entry flow: $(h+\frac{7}{6}a)\frac{d^2h}{dt^2}+1.225(\frac{dh}{dt})^2+\frac{8\mu}{\rho a^2} h\frac{dh}{dt}+g h=\frac{2\sigma}{\rho a}\cos \theta$.  Maggi, et al. \cite{mag} and Bush \cite{bus} derived a similar equation with a different coefficient of $(\frac{dh}{dt})^2$. The singularity-free equation of capillary rise dynamics was formulated as follows:
\begin{equation}\label{f3}
  (h+\frac{7}{6}a)\frac{d^2h}{dt^2}+\frac{1}{2}(\frac{dh}{dt})^2+\frac{8\mu}{\rho a^2} h\frac{dh}{dt}+g h=\frac{2\sigma}{\rho a}\cos \theta,
\end{equation}
with initial height and velocity boundary conditions of $h(0)=0, \, \dot{h}(0)=0$. (see the Supplementary).

Despite researching capillary rise dynamics for centuries, however, no satisfied complete solution has yet been found. Some obtained asymptotic solutions have never been matched together to form a complete solution that is valid for the whole time domain, therefore, capillary rise dynamics is still remained as unsolved problem. It would be natural attempts to seek a complete solution that is valid for entire time domain.

Eq.(\ref{f3}) is a nonlinear ordinary differential equation, whose series solution is $h(t)=\sum_{n=0}^\infty c_n t^n= \frac{6 \sigma \cos \theta}{7\rho a^2}t^2-\frac{3\sigma \cos \theta (7g \rho a^2+24\sigma \cos \theta)}{343\rho^2a^5}t^4-\frac{864\mu\sigma^2\cos^2 \theta}{1715\rho^3 a^7}t^5+O(t^6)$, which gives the non-zero initial acceleration $\ddot{h}(0)=\frac{12}{7} \frac{ \sigma \cos \theta}{\rho a^2}$. However, the rate of convergence of this solution is slow and useless for practical application. It would be natural to propose a simple solution for the capillary rise dynamics.

In physics, the entire capillary dynamics process can be qualitatively described as follows: in an infinite reservoir the capillary rise and the velocity are zero in the initial state. Due to the effect of the surface tension, the capillary liquid obtains initial acceleration (the initial acceleration must never be zero), and begins to rise at a relatively uniform velocity, while the surface tension plays a dominant role in the ascending phase; however, as the capillary rises, wall frictions and gravity begin to work in an attempt to prevent the rise of the capillaries, and their joint action succeeds in decelerating the capillaries to a point, until the capillaries are finally stopped. Surface tension and wall resistance, as well as gravity to achieve unity of opposites, and the capillary dynamics process are attributed to calm.

From the above perspective, the capillary liquid will rise to a stationary level, while the height and velocity result in $h \rightarrow H=\frac{2\sigma \cos \theta}{\rho g a}$ and $\dot{h} \rightarrow 0$ at $t \rightarrow \infty$, respectively. To satisfy the conditions, we can obtain the capillary rise (see the Supplementary materials)
\begin{equation}\label{eq-2}
  h(t)=H[1-(1+\beta t+\alpha t^3)e^{-\beta t}].
\end{equation}
Therefore, the capillary velocity $\dot{h}(t)=H(\alpha \beta t^2-3\alpha t+\beta^2)t e^{-\beta t}$ and acceleration $ \ddot{h}(t)=-H(\alpha \beta^2 t^3+\beta^3 t-6\alpha \beta t^2+6\alpha t-\beta^2)e^{-\beta t}$, where $\beta=\sqrt{\frac{6g}{7a}}$, and $\alpha \approx \frac{7\mu\sigma\cos\theta}{12\rho^2a^5}-\frac{1}{15}\left(\sqrt{\frac{6g}{7a}}\right)^3$. The solution in Eq.(\ref{eq-2}) is valid for the entire time domain $t\in [0,\infty)$, which, to the author's knowledge, has never been reported in literature.

The solution in Eq.(\ref{eq-2}) reveals that the capillary rise $h(t)$ is mainly controlled by the $H=\frac{2 \sigma \cos \theta}{\rho g a}$ with the decay rate of $e^{-\left(\sqrt{\frac{6g}{7a}}\right)t}$, where the radius $a$ is the only dominant parameter, and the smaller radius represents the fast decay.

For a validation and comparative study, the data of the diethyl ether in glass, is shown in the table below \ref{tb1}:

\begin{table}[h]
\caption{The diethyl ether in glass with arbitrary radius}\label{tb1}
\footnotesize
\centerline{
\begin{tabular}{cccccc}
\hline
\hline
$\mu\,[\frac{kg}{ms}]$ &  $\sigma \,[\frac{kg}{s^2}]$ &  $\theta$ & $g\,[\frac{m}{s^2}]$ & $\rho\,[\frac{kg}{m^3}]$\\
\hline
$2.2\cdot 10^{-4}$ &  $1.67\cdot 10^{-2} $ &  $26^0$ & $9.81$ & $710$\\
\hline
\hline
\end{tabular}}
\end{table}

In this case, for any radius $a$, we have the capillary rise $H=\frac{4.31}{a}\cdot10^{-6}$, hence the analytical capillary rise is:
\begin{equation}\label{aa}
\begin{split}
  h(t)&=\frac{4.31 \cdot 10^{-6}}{a}\{1-\\
 & [1+\frac{2.9}{\sqrt{a}}t+(\frac{3.821\cdot 10^{-12}}{a^5}-\frac{1.626}{a^{3/2}})t^3]e^{-\frac{2.9}{\sqrt{a}}t}\}.
  \end{split}
\end{equation}
In the case of tube radius $a=0.0005$, the capillary rise, the velocity and the acceleration are plotted in (a,b,c) of Figure 2. Regarding the influence of radius change, the Figure 2(d) shows that the capillary rise is sensitive to the tube radius. The smaller the radius in which the liquid can travel, the further it goes. Therefore, the glass tube radius is a crucial and domination parameter for capillary rise dynamics.

\begin{figure}[h!]
\centerline{\includegraphics[scale=0.3]{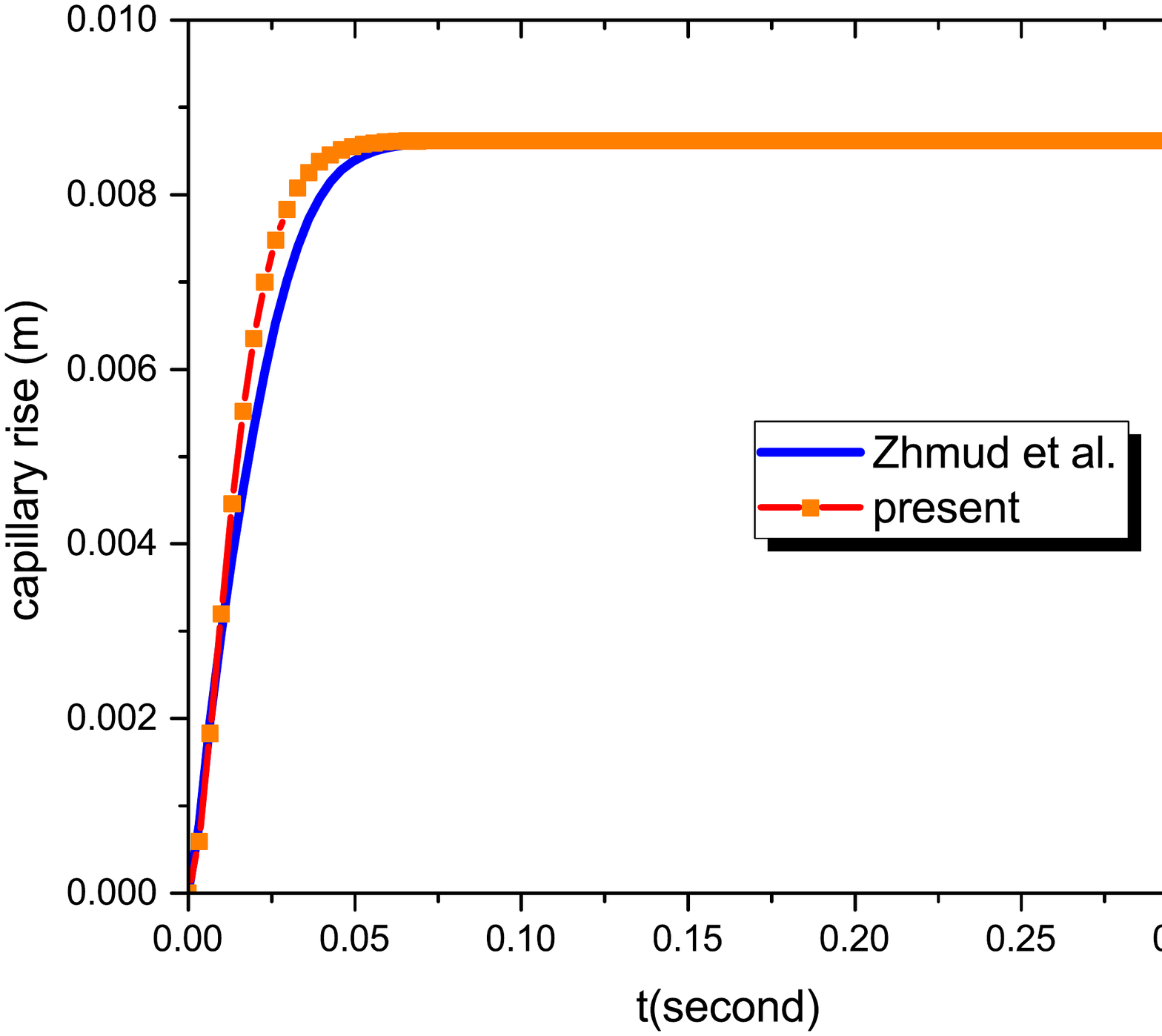}(a)}
\centerline{\includegraphics[scale=0.3]{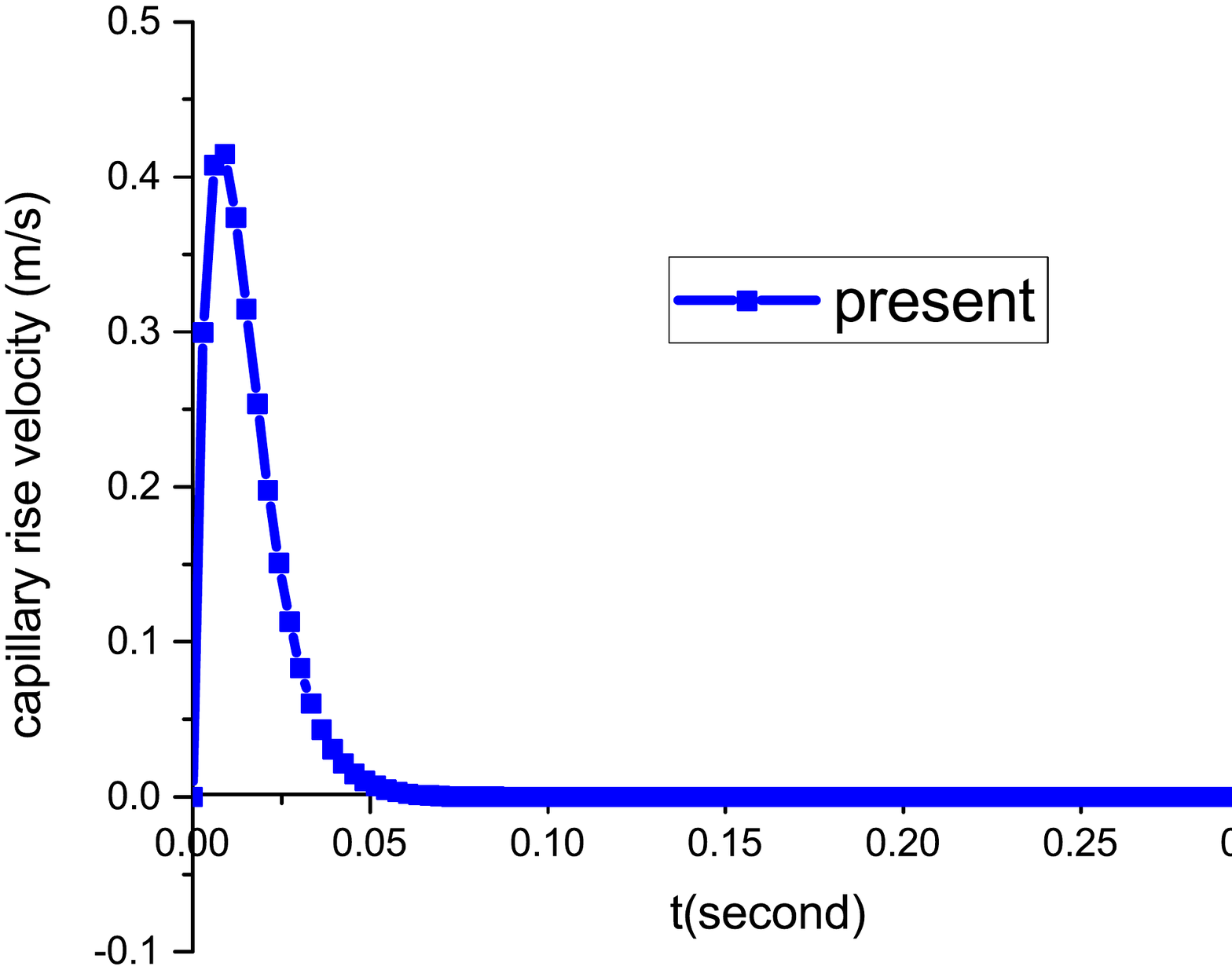}(b)}
\centerline{\includegraphics[scale=0.3]{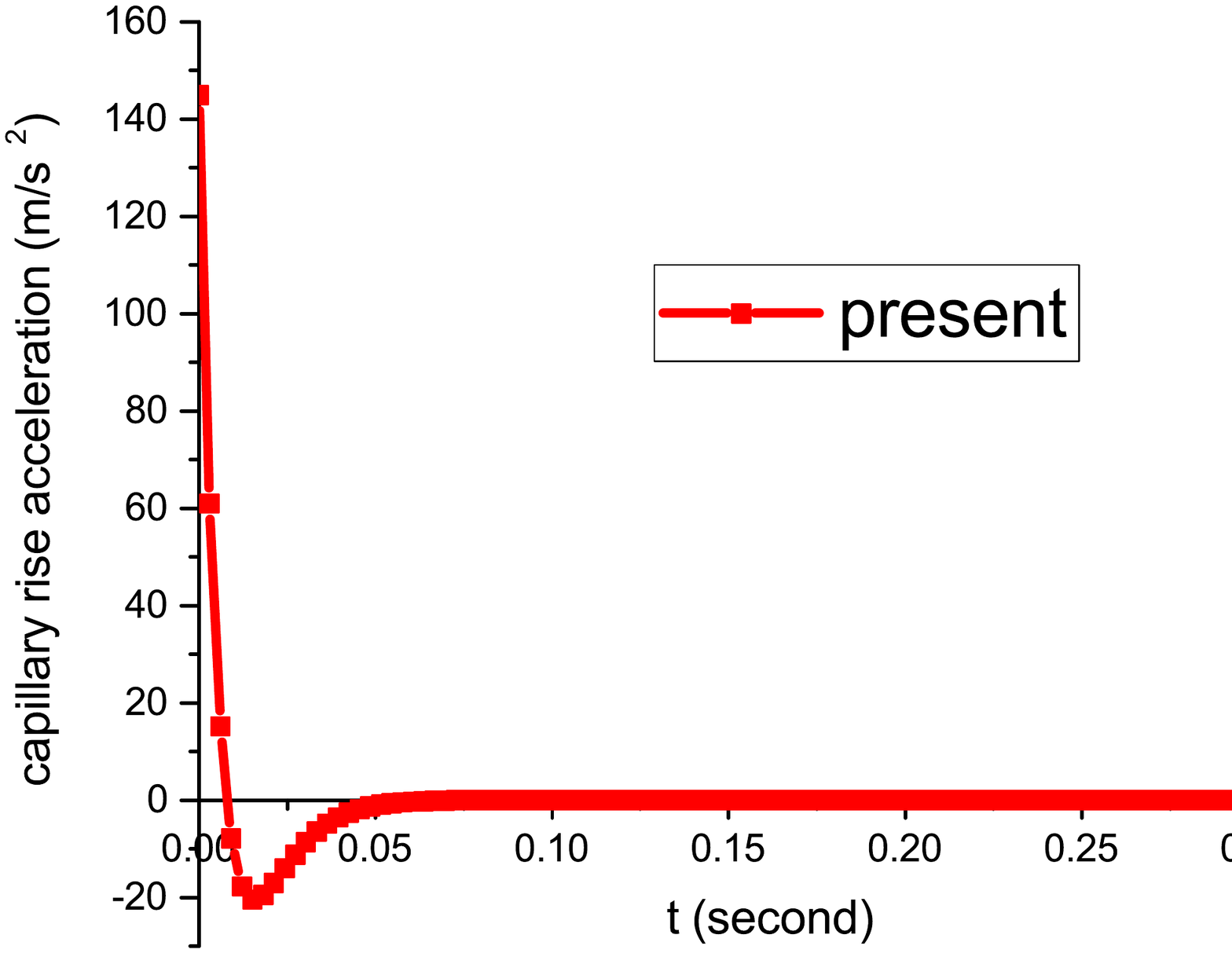}(c)}
\centerline{\includegraphics[scale=0.3]{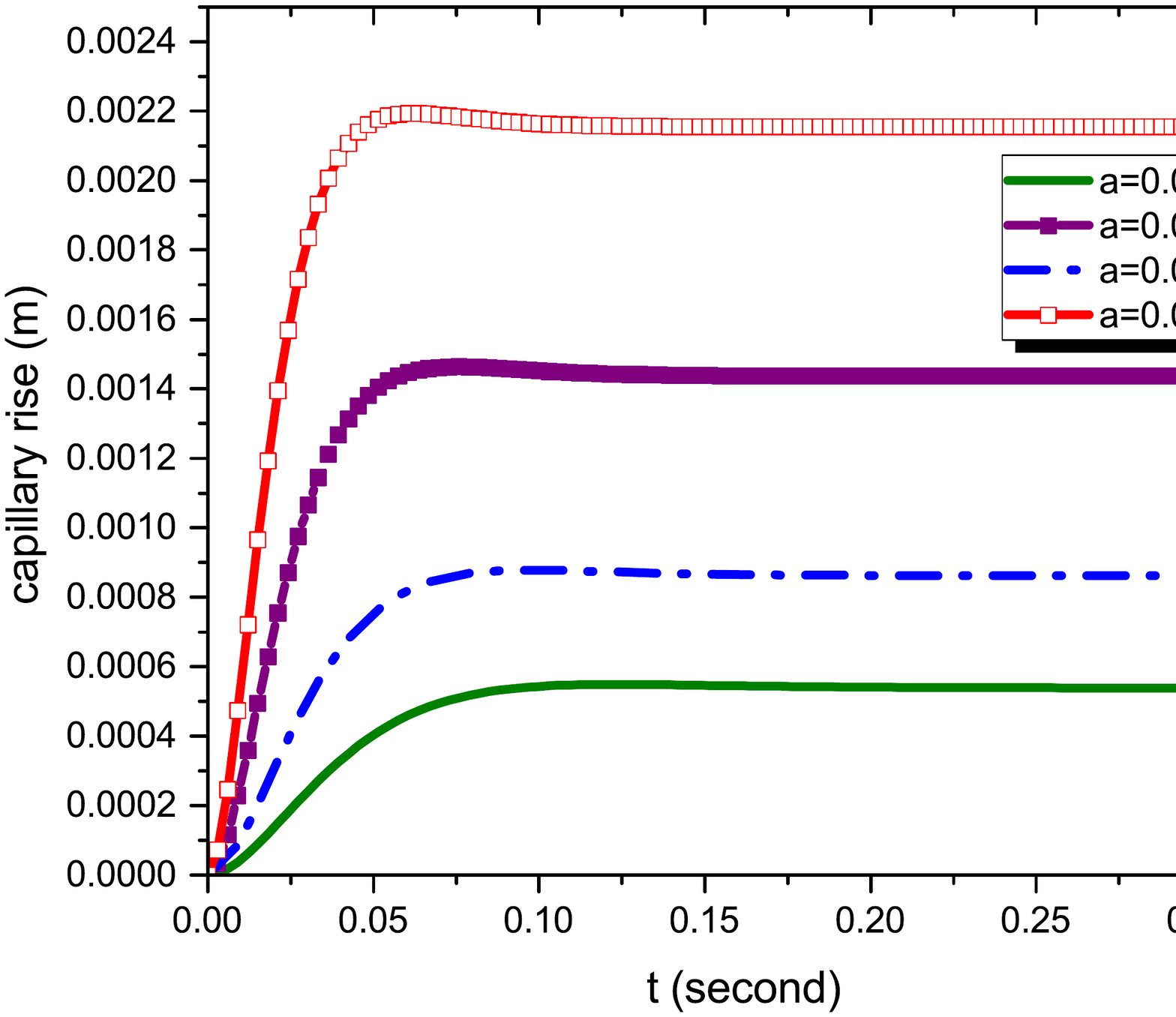}(d)}
\caption{\label{fig005} (a) The capillary rise $h(t)=0.00862\left[1-(1+129.68t-23112.54t^3)e^{-129.68t}\right]$ is fairly agree with Zhmud et al. \cite{zhm}; (b) the capillary velocity $\dot{h}(t)=-25836t(t+0.06)(t-0.08)e^{-129.68t}$; (c) the capillary acceleration $\ddot{h}(t)=3350496(t+0.006)(t-0.008)(t-0.096)e^{-129.68t}$, (d) four different tube radius comparison, which shows that the tube radius is domination parameter in the dynamics.}
\end{figure}

In conclusion, from Figure 2, the capillary rise dynamics can be interpreted as follows: the capillary rise onsets from an initial zero height and velocity at $t=0$, the surface tension, which is the first driving force that supplies a kick-off acceleration at $H\beta^2$. This surface tension steadily drives the capillary rising at an almost uniform speed until it reaches its peak at a certain point $H$; then the capillary velocity $\dot{h}(t)$ is gradually reduced to zero after short oscillation owing to the combined effect of the wall viscous friction and gravity. At the same time, the capillary acceleration $\ddot{h}(t)$ decreases and reaches its lowest point before it increases to its final static state.

Last, but not least, it may be worth pointing out that all previous solutions were derived from the non-oscillatory regime, however, when the liquid surface reaches $h=H$, the surface will oscillate before asymptotically ending. Within the oscillatory regime, the above solutions can be modified by using perturbation method, namely, by setting $h(t)=H[1+\tilde{h}(t)]$, where $\tilde{h(t)} $ is a small perturbation \cite{zho}.

I am grateful to Prof. Shijun Liao, Yapu Zhao and Michael Sun for their interest and high level academic comments from reviewers.

\nocite{*}


\end{document}